\documentclass{revtex4-1}
\usepackage{amsmath}
\usepackage{bm}
\usepackage{titlesec}
\usepackage{caption}
\usepackage{subcaption}
\usepackage{tikz}
\titlespacing\subsection{0pt}{6pt plus 4pt minus 2pt}{0pt plus 2pt minus 2pt}

\newcommand{\ve}{\vec{E}}
\newcommand{\vb}{\vec{B}}
\newcommand{\vu}{\vec{u}}
\newcommand{\veps}{\vec{\epsilon}}

\newcommand{\exb}{{\vec{E} \times \vec{B}}}

\newcommand{\vbet}{\vec{\beta}}

\begin{document}
\title{Structure-preserving second-order integration of relativistic charged particle trajectories
in electromagnetic fields}
\author{A.V. Higuera}
\email{adam.higuera@colorado.edu}
\affiliation{University of Colorado at Boulder}
\affiliation{Tech-X Corporation}
\author{J.R. Cary}
\email{cary@txcorp.com}
\affiliation{University of Colorado at Boulder}
\affiliation{Tech-X Corporation}

\begin{abstract}


Time-centered, hence second-order, methods for integrating the
relativistic momentum of charged particles in an electromagnetic
field are derived.  A new method is found by averaging the
momentum before use in the magnetic rotation term, and an
implementation is presented that differs from the relativistic
Boris Push~\cite{boris1970relativistic} only in the method for calculating the Lorentz factor.
This is shown to have the same second-order accuracy in time as
that (Boris Push) \cite{boris1970relativistic} found by splitting
the electric acceleration and magnetic rotation and that
\cite{vay2008simulation} found by averaging the velocity in the
magnetic rotation term.  All three methods are shown to conserve
energy when there is no electric field.  The Boris method and the
current method are shown to be volume-preserving, while the method of
\cite{vay2008simulation} and the current method preserve the
$\exb$ velocity.  Thus, of these second-order relativistic
momentum integrations, only the integrator introduced here both
preserves volume and gives the correct $\exb$ velocity.  While
all methods have error that is second-order in time, they deviate
from each other by terms that increase as the motion becomes
relativistic.  Numerical results show
that~\cite{vay2008simulation} develops energy errors near
resonant orbits of a test problem that neither volume-preserving
integrator does.

\end{abstract}

\maketitle

\section{Introduction}

Computing trajectories of classical charged particles
in an electromagnetic field is critical to many areas of physics.
In beam physics (cf~\cite{cary2016select} and references therein)
particle tracking through accelerator lattices is needed to
determine whether particle will remain in, e.g., storage rings
for times long enough for significant reactions to occur.  In
self-consistent plasma physics computations, the computation of
particle trajectories is one part of kinetic modeling of plasmas
using the Particle-In-Cell (PIC)
method~\cite{birdsall2004plasma,hockney1988computer,nieter2004vorpal}.
Particle tracking is also important for understanding the limits
of precision in spectrometers.

The underlying system is Hamiltonian, and so has a complete set
of Poincar\'e geometric invariants~\cite{goldstein2014classical}.
Symplectic integrators are those that also preserve the
Poincar\'e invariants.  Symplectic integrators allow one to
invoke the KAM theorem, so that the integrated motion shadows the
real motion (except for the very slow Arnold diffusion in systems
with 2.5 or more degrees of freedom).  Explicit symplectic
integrators~\cite{ruth1983canonical,feng1986difference,forest1990fourth,candy1991symplectic,yoshida1990construction}
relevant to particle tracking are now in common use, and there
are also explicit symplectic integrators~\cite{cary1993explicit}
for certain plasma systems.  Unfortunately, there is at present
no known explicit symplectic integrator for charged particle
motion in arbitrary electromagnetic fields.

A less restrictive condition is that the integrator be
volume-preserving, i.e., that it preserve the last Poincar\'e
invariant.  This prevents attractors and repellers in the
integrated system, important as they do not exist in the
underlying system.  The relativistic Boris
integrator~\cite{boris1970relativistic} (also known as the
\emph{Boris push}) is volume-preserving (more detailed discussion
in light of \cite{qin2013boris,zhang2015volume} below) and is
found to have good properties when used to compute trajectories.
In particular, the spatial Boris push~\cite{stoltz2002efficiency}
has been found excellent for tracking studies in accelerators, as
it eliminates numerical cooling or heating, which can mask
physical effects due to small dissipative terms.

On the other hand, the relativistic Boris push does not correctly
compute the $\exb$ drift velocity, and this was
found~\cite{vay2008simulation} to lead to problems in some
special cases.  Vay proposed using a different integrator that
does preserve the $\exb$ drift velocity at the cost of a slightly
more complicated computation of the new relativistic factor.
Unfortunately, Vay's integrator is not volume-preserving, as we
show here, and so it lacks an important feature of the Boris
integrator.  The lack of volume-preservation is shown to lead to
unphysical behavior in the computed trajectories.

Here we construct an integrator that preserves both the $\exb$
drift velocity and phase-space volume.  Like both the Boris and
Vay integrators, it also conserves energy in the absence of an
electric field.  This new integrator has computational cost
comparable to that of the Vay integrator.

The organization of this paper is as follows.  In the following
section we introduce the leap-frog separation of the spatial and
momentum integrations that reduce this to second-order
integration of the momentum change equation.  In the same section we
discuss also the various centerings that can be used obtain a
second-order-accurate momentum integration, deriving a new
integrator and showing that the previous Boris and Vay
integrators follow from this principle and have the same order of
accuracy.  Section~\ref{explicit} discusses implementation.  In Sec. \ref{preserve} we show that all three integrators
conserve energy when there is no electric field, but that only
the current integrator and the integrator of
\cite{vay2008simulation} properly capture the $\exb$ drift.  In
Sec.~\ref{volumepres} we show that only the current and Boris
integrators are volume-preserving.  In Sec.~\ref{numerics} we
show that the new and Boris integrators eliminate un-physical
behavior displayed by the integrator of~\cite{vay2008simulation}
in a test problem.
Finally, in Sec.~\ref{summfuture} we summarize our results and
conclude.

\section{Second-order charged particle integrators}
\label{secondorderint}


Charged particles in the electromagnetic field follow the
dynamics of relativistic translation,
\begin{equation} \label{eq:reltrans}
  \frac{d \vec{x}}{dt} = \vec{v}(\vec{u}) = \vec{u}/\gamma,
\end{equation}
where $\gamma = \sqrt{1 + |\vec{u}/c|^2}$, and the
Lorentz force,
\begin{equation} \label{eq:lorentz}
  \frac{d \vec{u}}{dt} = (q/m)(\vec{E} + \vec{v}(\vec{u}) \times \vec{B}).
\end{equation}
The first step to an efficient, second-order integration of these
equations is use of the Leap-Frog method, in which the
finite-time-step relativistic translation becomes
\begin{align} \label{eq:reltransdt}
  \vec{x}_f &= \vec{x}_i + \Delta \vec{x}\text{ with} \\
  \Delta \vec{x} &= \vec{v} \Delta t,
\end{align}
and then one is left with solving the Lorentz force equation
Eq.~(\ref{eq:lorentz}) to second order for a time step,
$\Delta t$, with the assumption of constant electric and magnetic
fields.


Ideally one would like to solve this equation in a way that
preserves as many properties of the underlying differential
equations as possible.  In this paper we consider the following
properties:  (1)~Energy should be conserved in the absence of an
electric field.  (2)~The static solution for crossed electric and
magnetic fields, with $|\vec{E}| < c|\vec{B}|$, should be
constant velocity in the third direction of magnitude
$|\vec{E}|/|\vec{B}|$.  (3)~The differential volume, which is
preserved by any solution of the differential equation, should be
preserved by the finite-time-step solution.

The standard way to obtain a second-order solution is by time
centering.  That is, one uses the solution,
\begin{align}
  \vec{u}_f &= \vec{u}_i + \Delta \vec{u}\text{ with} \label{eq:lorentzdta} \\
  \Delta \vec{u} &= (q/m)(\vec{E} + \bar{v} \times \vec{B}) \Delta t, \label{eq:lorentzdtb}
\end{align}
where $\bar{v}$ is an average of the initial and final values of
$\vec{v}$.  There are multiple choices for how to do the average.
Here we introduce a new choice,
\begin{equation} \label{eq:higuera-cary}
  \bar{v}_{new} \equiv \vec{v}\left(\frac{\vec{u}_i + \vec{u}_f}{2}\right).
\end{equation}
while \cite{vay2008simulation} made the choice,
\begin{equation} \label{eq:vay}
  \bar{v}_v \equiv \frac{\vec{v}(\vec{u}_i) + \vec{v}(\vec{u}_f)}{2}
\end{equation}
and the Boris choice is,
\begin{equation} \label{eq:boris}
  \bar{v}_b \equiv \frac{\vec{v}\left(u_i + \vec{\epsilon}\right) + \vec{v}\left(u_f - \vec{\epsilon}\right)}{2}.
\end{equation}
where
\begin{equation} \label{eq:accel}
  \vec{\epsilon} \equiv \frac{q \vec{E}}{2m} \Delta t.
\end{equation}
The Boris push is not usually written in this way, but one can
see that Eq.~(\ref{eq:boris}) is equivalent to a magnetic
rotation by the velocity found after an initial half electric
acceleration (the first term in the numerator) and before a
final half acceleration (the second term in the numerator).

All of these integrators have second-order error in the time step, as
one can see from Taylor expansion.  E.g.,
\begin{align} \label{eq:vayexp}
  \bar{v}_v &= \frac{\vec{v}(\bar{u}_v + \Delta \vec{u}/2)}{2} + \frac{\vec{v}(\bar{u}_v - \Delta \vec{u}/2)}{2} =
  \vec{v}(\bar{u}_v) + O(\Delta t^2),
\end{align}
is equivalent to Eq.~(\ref{eq:higuera-cary}) to second order.
That the Boris push has the same order of error follows from a
similar calculation.

\section{Explicit evaluation}
\label{explicit}

The new integrator (\ref{eq:lorentzdta}-\ref{eq:higuera-cary})
looks implicit, as the final momentum on the left side of
(\ref{eq:lorentzdta}) is involved in its definition through
(\ref{eq:higuera-cary}).  However, it can be explicitly computed
by methods similar to that of \cite{vay2008simulation}.   We
first write the new integrator as the composition of two
integrators.
\begin{align}
  \vec{u}_f &= \bar{u}_{new} + \Delta \vec{u}_{new}/2\text{ and} \label{eq:finalhchalfstep} \\
  \vec{u}_{new} &= \bar{u}_i + \Delta \vec{u}_{new}/2 \label{eq:inithchalfstep}\text{ or} \\
  \vec{u}_i &= \bar{u}_{new} - \Delta \vec{u}_{new}/2 \label{eq:inithchalfstep2},
\end{align}
where
\begin{equation}
  \Delta \vec{u}_{new} = 2 \vec{\epsilon} +
  \frac{\bar{u}_{new}}{\gamma_{new}} \times 2 \vec{\beta}, \label{eq:duhcdef}
\end{equation}
with
\begin{equation}
  \vec{\beta} \equiv \frac{q \vb}{2 m} \Delta t, \label{eq:betadef}
\end{equation}
and $\gamma_{new} \equiv \gamma(\vec{u}_{new})$.  The first
equation (\ref{eq:finalhchalfstep}) corresponding to the last
half update is explicit.  So we need only solve
Eqs.~(\ref{eq:inithchalfstep2}-\ref{eq:betadef}) to obtain an
explicit solution.

In a process similar to that of \cite{vay2008simulation},
the explicit solution comes from taking the dot and cross products of
Eq.~(\ref{eq:inithchalfstep2}) and clearing terms to obtain
\begin{equation}
  \bar{u}_{new}\left(1 + \frac{\beta^2}{\gamma_{new}^2}\right) = \vec{u}_- -
      \frac{\vec{\beta} \times \vec{u}_-}{\gamma_{new}} + \frac{\vec{\beta}
      \vec{\beta} \cdot \vec{u}_-}{\gamma_{new}^2},  \label{eq:ubargammaimpl}
\end{equation}
in which
\begin{equation}
  \vec{u}_- \equiv \vec{u}_i + \vec{\epsilon}. \label{eq:uminusdef}
\end{equation}
Eq.~(\ref{eq:ubargammaimpl}) gives $\bar{u}_{new}$ explicitly provided $\gamma_{new}$ can be computed.  Squaring Eq.~(\ref{eq:ubargammaimpl}) gives the
biquadratic polynomial,
\begin{equation}
  (\gamma_{new}^2 - 1)\left(\gamma_{new}^2 + \beta^2\right) =
       \gamma_{new}^2 |\vec{u}_-|^2 + |\vec{\beta} \cdot \vec{u}_-|^2
\end{equation}
which can be solved to obtain
\begin{equation}
  \label{eq:gamma_condition}
  \gamma_{new}^2 = \frac{1}{2} \left(\gamma_-^2 - \beta^2 +
      \sqrt{(\gamma_-^2 - \beta^2)^2 + 4(\beta^2 + |\vec{\beta} \cdot \vec{u}_-^2|) } \right),
\end{equation}
where
\begin{equation}
  \gamma_- \equiv \gamma(\vec{u}_-) \label{eq:gammaminusdef}.
\end{equation}
Equation~(\ref{eq:ubargammaimpl}) is needed to derive
Eq.~(\ref{eq:gamma_condition}) for
$\gamma_{new}$, but, once $\gamma_{new}$ is known, $\bar{u}_{new}$ is not necessary
to obtain $\vu_f$.  Using the definition of $\Delta \vu_{new}$ and
Eqs.~(\ref{eq:finalhchalfstep}-\ref{eq:inithchalfstep2})
\begin{align}
  \label{eq:expl_steps}
  \vu_i + \veps &= \vu_{new} - \vu_{new} \times
                            \frac{\vbet}{\gamma_{new}} \\
  \vu_f - \veps &= \vu_{new} + \vu_{new} \times \frac{\vbet}{\gamma_{new}}\label{eq:expl_steps2}.
\end{align}
Defining then and $\vu_{+} = \vu_{f} -
\veps$, using Eq.~(\ref{eq:uminusdef}), and subtracting Eq.~(\ref{eq:expl_steps}) from
Eq.~(\ref{eq:expl_steps2}) yields the familiar Boris rotation
equation:
\begin{equation}
  \label{eq:boris_rot_eq}
  \vu_+ - \vu_- = (\vu_+ + \vu_-) \times \frac{\vbet}{\gamma_{new}}
\end{equation}

The new integrator's implementation differs from the Boris
integrator's in the use of Eq.~(\ref{eq:gamma_condition})
instead of Eq.~(\ref{eq:gammaminusdef}) to calculate $\gamma$ for
the magnetic rotation.  The two prescriptions differ by terms
second-order in $\Omega_c \Delta t$.  One can furthermore
show that the relativistic factor used by the Boris integrator
always exceeds or equals that of the present integrator.
\begin{equation}
  \gamma_- \ge \gamma_{new} \label{eq:gammainequality}
\end{equation}

\section{Preservation of limiting solutions}
\label{preserve}

Here we analyze the above integration methods with regard to
whether they preserve properties of limiting solutions, (1)~no
electric field and (2)~correct value of perpendicular velocity in
crossed electric and magnetic fields.  In the first case, the
differential equations have no change of energy.  This is well-known to be true for the Boris push.  It follows that it is true
for the other integrators, as, without an electric field, the
relativistic factor is preserved even for the finite difference
solution, and so both the new integrator~(\ref{eq:higuera-cary})
and the Vay integrator~(\ref{eq:vay}) preserve energy when there
is no electric field.

When there is no electric field and the magnetic
field is static and uniform, the Boris integrator rotates the particle's
momentum through an angle
\begin{equation}
  \label{eq:boris_rot_angle}
  \theta = \Omega_c \gamma^{-1} \Delta t \left(1 - \frac{(\Omega_c \gamma^{-1}
      \Delta t)^2}{12}  + \mathcal{O}((\Omega_c \gamma^{-1} \Delta t)^4) \right)
\end{equation}
It can be shown that, for the same problem, the new integrator rotates
the particle's momentum through an angle
\begin{equation}
  \label{eq:hc_rot_angle}
  \theta = \Omega_c \gamma^{-1} \Delta t \left(1 + \left[ \frac{1}{8} \left(
      1 - \frac{1}{\gamma^2} \right) - \frac{1}{12} \right] (\Omega_c \gamma^{-1}
  \Delta t)^2 + \mathcal{O}((\Omega_c \gamma^{-1} \Delta t)^4) \right).
\end{equation}
In the non-relativistic case, Eq.~(\ref{eq:hc_rot_angle}) reduces to
Eq.~(\ref{eq:boris_rot_angle}), with the coefficient of the
third-order error term increasing monotonically to a limit of $+1/24$
in the ultra-relativistic case (vanishing at $\gamma = \sqrt{3}$).
The new integrator's third-order error term is therefore always
smaller than the Boris integrator's for electron gyro-motion with
$\Omega_c \Delta t \ll 1$.

For perpendicular electric and magnetic fields, a stationary
solution, $\Delta \vec{v} = 0$, must have, from
Eq.~(\ref{eq:lorentzdtb}),
\begin{equation}
  \bar{v} = \vec{E} \times \vec{B}/|\vec{B}|^2.
\end{equation}
Stationarity also implies $\vec{u}_f = \vec{u}_i$, which used
in either of Eqs.~(\ref{eq:higuera-cary},\ref{eq:vay}) implies
that
\begin{equation}
  \bar{v}_f = \bar{v}_i = \vec{E} \times \vec{B}/|\vec{B}|^2.
\end{equation}
Hence for either the new integrator or the Vay integrator, the
electric drift velocity has the same value as for the
differential equations.  For the Boris integrator, as noted in
\cite{vay2008simulation}, this is not the case.
Inequality~\ref{eq:gammainequality} explains why: the Boris integrator
always rotates the momentum less than the new integrator, which is
shown here to rotate $\vu_{-}$ exactly enough that the second
half-acceleration exactly cancels the first.

\section{Volume-preservation}
\label{volumepres}

It has been surmised~\cite{qin2013boris} that the excellent
conservation properties of the nonrelativistic Boris push, as,
e.g., observed in \cite{stoltz2002efficiency}, are due to the
fact that it preserves phase space volume, like the underlying
differential equations.  This follows straightforwardly from the
observation that the nonrelativistic Boris push is a sequence of
sheared (spatial change) translations, unsheared (electric
acceleration) translations, and rotations (magnetic
acceleration).  An additional observation, that the relativistic
Boris push simply changes the rotation to being sheared (due to
the dependence of the rotation on the relativistic factor,
$\gamma$, which is constant for that part of the transformation)
shows that the relativistic Boris push is also volume-preserving.
(The reference \cite{zhang2015volume} shows that an integrator
that is not actually the Boris integrator does not preserve volume
but then ultimately gives a more detailed proof of volume
preservation by what is in fact the Boris integrator.)

While the two transformations
(\ref{eq:finalhchalfstep},\ref{eq:inithchalfstep}) are not
inverses of each other (otherwise the integrator would be the
identity), we will show that their two Jacobian determinants are reciprocals of
one another and so the transformation for the new integrator is
volume-preserving.


Differentiating (\ref{eq:finalhchalfstep}) gives the Jacobi matrix,
\begin{equation}
  \frac{\partial (\vec{u}_f)}{\partial (\bar{u}_{new})} = \bm{I} - \bm{\Omega} + \frac{\vec{\beta}_{new} \times \bar{u}_{new} \otimes \bar{u}_{new}}{\gamma^3_{new}},
\end{equation}
where $\bm{\Omega}$ is the matrix defined by
\begin{equation}
  \bm{\Omega} \cdot \vec{V} \equiv \vec{\beta} \times \vec{V}/\gamma_{new}.
\end{equation}
By the determinant lemma, this matrix has determinant,
\begin{equation}
  J_{f,new} \equiv \det\left(\frac{\partial (\vec{u}_f)}{\partial (\bar{u}_{new})}\right) = \det(\bm{I} - \bm{\Omega}) \det\left(1 + \frac{\bar{u}_{new} \cdot (\bm{I} - \bm{\Omega})^{-1} \cdot (\vec{\beta} \times \bar{u}_{new})}{\gamma^3_{new}}\right). \label{eq:endjac}
\end{equation}
Explicit computation gives
\begin{equation}
  \det(\bm{I} - \bm{\Omega}) = 1 + \beta^2/\gamma_{new}^2 \label{eq:detoneminusbeta_{new}}.
\end{equation}
Explicit computation can also be used to find
\begin{equation}
  \bar{u}_{new} \cdot (\bm{I} - \bm{\Omega})^{-1} \cdot (\vec{\beta} \times \bar{u}_{new}) = ((\bm{I} + \bm{\Omega})^{-1} \cdot \bar{u}_{new}) \cdot (\vec{\beta} \times \bar{u}_{new}) \label{eq:numeratorfactor}.
\end{equation}
by noting that
\begin{align}
  \vec{Y} &= (\bm{I} + \bm{\Omega})^{-1} \cdot \bar{u}_{new} \text{ or} \\
  \bar{u}_{new} &= \vec{Y} + \vec{\beta} \times \vec{Y}/\gamma_{new}
\end{align}
can be solved to give
\begin{equation}
  \vec{Y} = \frac{\vec{u}_{new} - \vec{\beta} \times \vec{u}_{new}/\gamma_{new} + \vec{\beta} \vec{\beta} \cdot \vec{u}_{new}/\gamma_{new}^2}{1 + \beta^2/\gamma_{new}^2},
\end{equation}
which can be used in Eq.~(\ref{eq:numeratorfactor}) and ultimately with
Eqs.~(\ref{eq:endjac} and \ref{eq:detoneminusbeta_{new}}) to obtain
\begin{align}
  J_{f,new} &= 1 + \frac{\beta^2 + (\vec{\beta} \cdot \bar{u}_{new})^2}{\gamma_{new}^4}.
\end{align}
By an identical process, one can show that
\begin{equation}
  J_{i,new} = 1 + \frac{\beta^2 + (\vec{\beta} \cdot \bar{u}_{new})^2}{\gamma_{new}^4}.
\end{equation}
Thus, the Jacobian of the first half step equals the inverse of
the Jacobian for the second half step, and so their product is
unity, and the new integrator is volume-preserving, just like the
Boris integrator.

The Vay integrator can be analyzed in this same way. It can likewise
be composed into two half steps, but with the first being an explicit
step using $u_i/\gamma_i$ followed by an implicit step using
$u_f/\gamma_f$.  As a result the full Jacobian for the Vay integrator
is
\begin{equation}
\label{eq:vay_jac}
  J_{v} = \frac{J(x_i, u_i)}{J(x_i,u_f)},
\end{equation}
where
\begin{align}
  J(x, u) &= 1 + \frac{\beta^2 + (\vec{\beta} \cdot \vec{u})^2}{\gamma^4}. \label{eq:genjac}.
\end{align}
Consequently, after N steps, the differential volume element in
the Vay integrator is
\begin{equation}
\label{eq:vay_jacprod}
  J_{v,B} = \frac{J(x_0, u_0)}{J(x_0,u_1)} \frac{J(x_1, u_1)}{J(x_1,u_2)} ... \frac{J(x_{N-1}, u_{N-1})}{J(x_{N-1},u_N)},
\end{equation}

Because the two Jacobians in any of the fractions of
Eq.~(\ref{eq:vay_jacprod}) depend on different variables (the initial
and final momenta) while having the same functional form, the Vay
integrator is not generally volume-preserving.  At subsequent
steps, the differential volume grows or shrinks by a similar
factor, but now evaluated at the new position.  The function,
$J(x, u)$ varies over space but is bounded for bounded regions.
Hence, generally the factors in Eq.~(\ref{eq:vay_jacprod}) are
variously greater and less than unity.  While one could imagine a
trajectory that conspires to have all factors either greater or
less than unity, we have not so far been able to construct such a
case.  In the case where the magnetic field is constant in space
and time, the series telescopes and then the boundedness of $J(x,
u)$ prevents the existence of attractors or repellers.

\section{Numerical results}
\label{numerics}

We choose a test problem of the form,
\begin{align}
  \label{eq:gen_test_problem}
  \ve &= E_x(x) \hat{x} \\
  \vb &= B_x(y) \hat{x},
\end{align}
for which
\begin{equation}
  \vec{A} = A_z(y) \hat{z}.
\end{equation}
The Hamiltonian for this system is
\begin{equation}
\label{eq:hamiltonian}
  H = \sqrt{1 + p_x^2 + p_y^2 + (p_z - A_z(y))^2} + \phi(x)
\end{equation}
in units of $q = m = c = 1$  This Hamiltonian is independent of
$z$, so $p_z$ is an invariant of the motion.  It is also time
independent, so $H$ is an invariant.  Further,
\begin{equation}
\label{eq:y-invariant}
  I_y \equiv p_y^2 + (p_z - A_z(y))^2
\end{equation}
is an invariant of the motion and in involution with the other
invariants.  Hence, this system of three degrees of freedom has
three invariants in involution and so is integrable.

To determine the degree to which the various integrators preserve
these invariants, we use the Poincar\'e surface of section
technique.  We follow the trajectories for initial conditions of
the same energy and same $p_z$ and plot the points $(y, p_y)$ in
the plane $x=0$ when that plane is crossed with positive $p_x$.
(This does require interpolation to that plane, when that plane
is crossed in some time step.)  We can then check how well the
invariant (\ref{eq:y-invariant}) is preserved by the integration.


For numerical integration, we have to be more specific.  We
choose $p_z = 0$ without loss of generality, as a different value
of $p_z$ is equivalent to choosing a different function $A_z(y)$.
We then choose the particular fields,
\begin{align}
  \label{eq:test_problem}
  \ve &= -a x \hat{x} \\
  \vb &= b y \hat{x},
\end{align}
which corresponds to a sheared magnetic field with the reversal
at $y = 0$.  For this system,
\begin{align}
  A_z(y)  &= \frac{1}{2} b y^2\\
  \phi (x) &= \frac{1}{2} a x^2.
\end{align}
We choose the energy $H = 4.$ to be moderately relativistic to expose
the effects of the varying volume element (\ref{eq:vay_jac}), and
we choose $a = 1$ and $b = 2$, i.e., of order unity.

\begin{figure}[htbp]
  \centering
  \begin{subfigure}[b]{0.3\textwidth}
    \includegraphics[width=\textwidth]{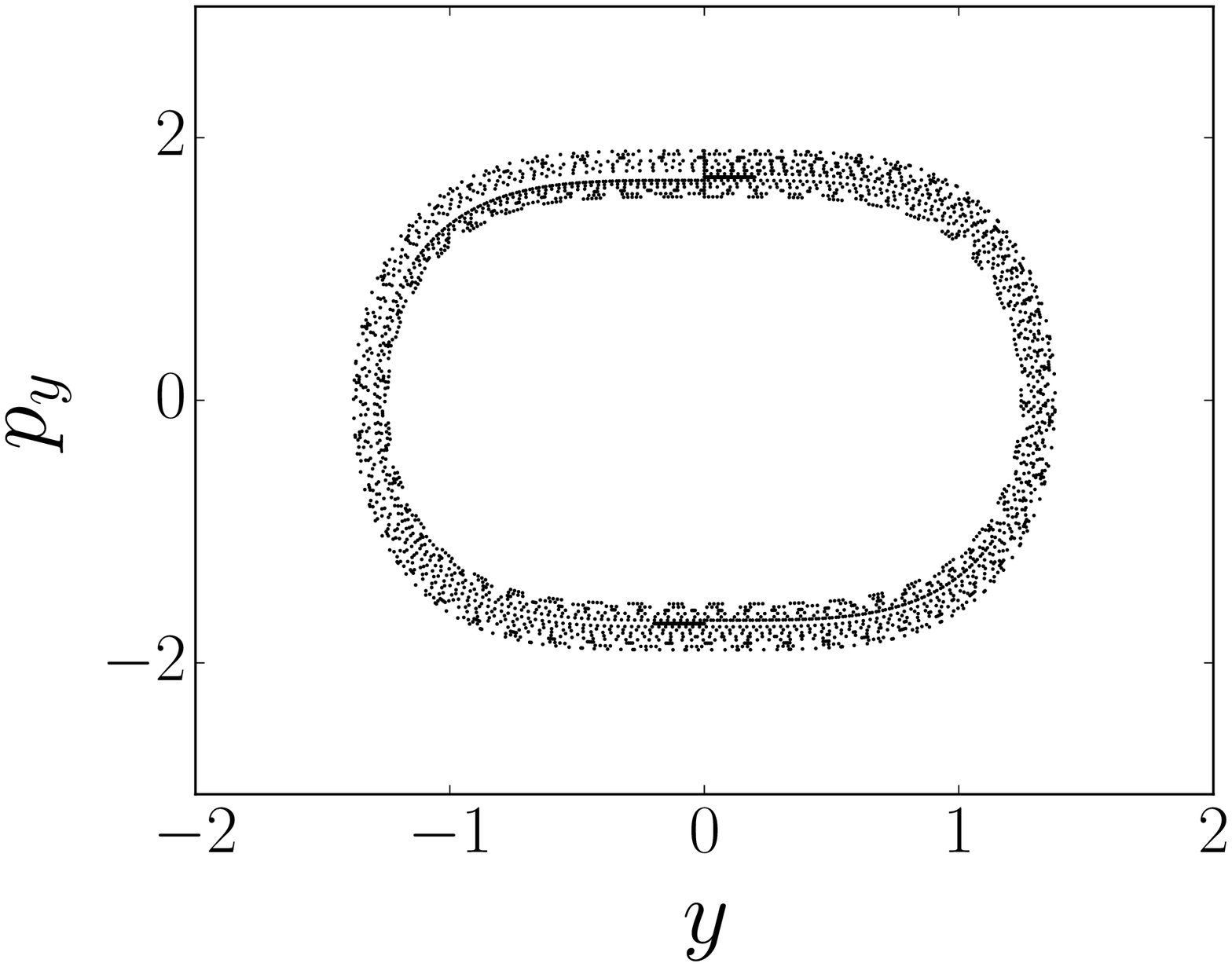}
    \caption{Boris}
    \label{fig:boris_sos-40}
  \end{subfigure}
  \begin{subfigure}[b]{0.3\textwidth}
    \includegraphics[width=\textwidth]{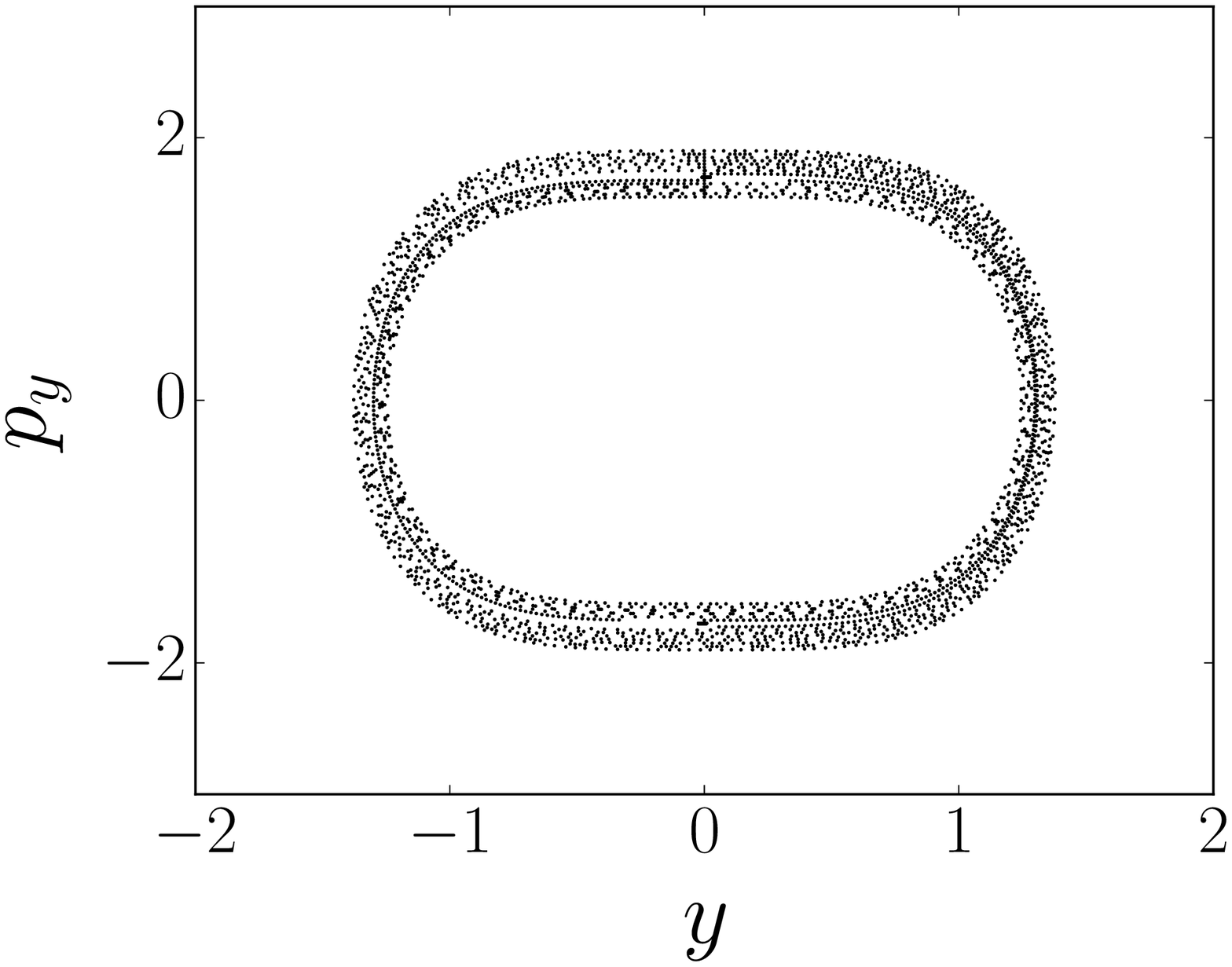}
    \caption{Vay}
    \label{fig:vay_sos-40}
  \end{subfigure}
  \begin{subfigure}[b]{0.3\textwidth}
    \includegraphics[width=\textwidth]{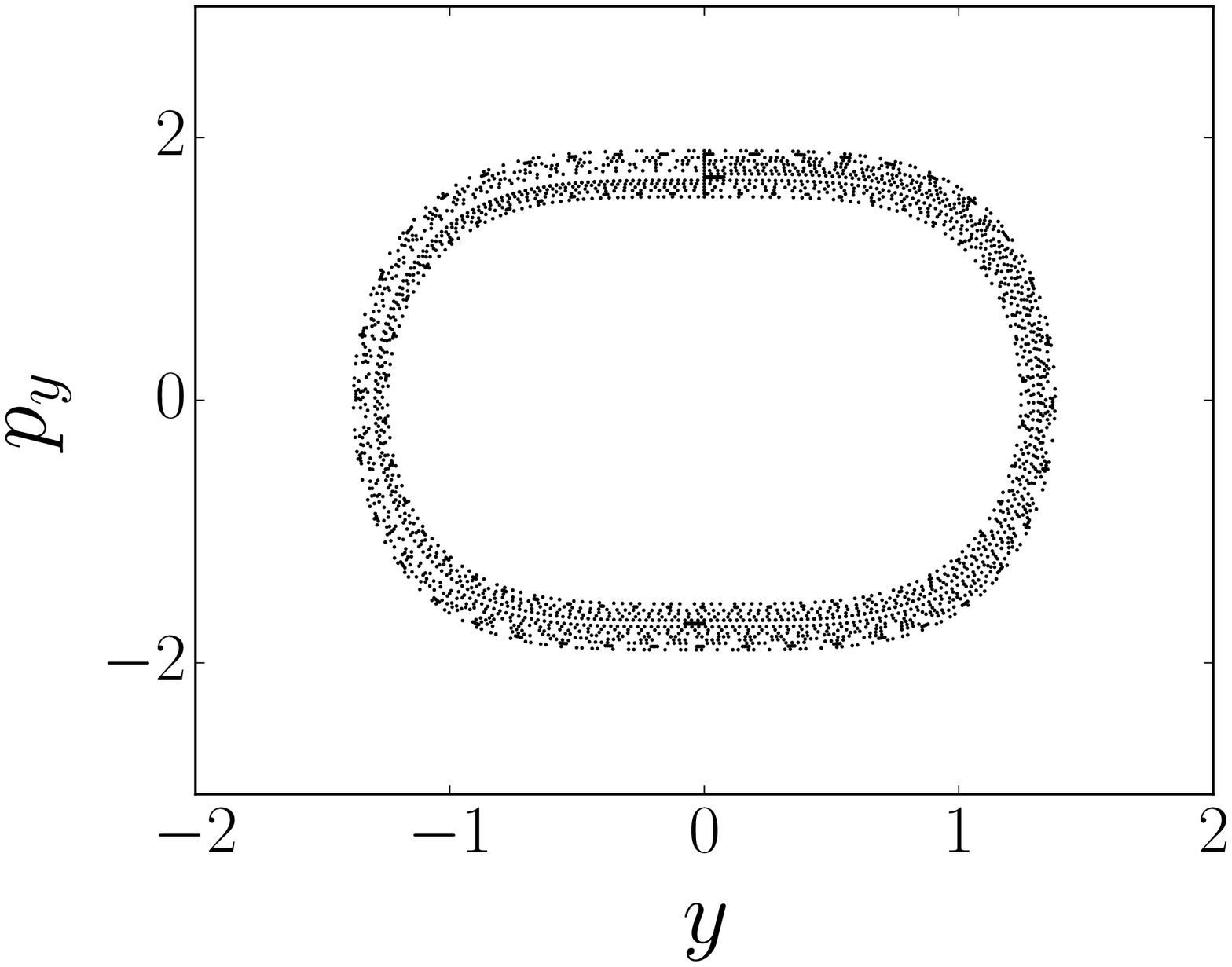}
    \caption{New}
    \label{fig:hc_sos-40}
  \end{subfigure}
  \caption{Poincar\'e surfaces of section for
    the system~(\ref{eq:test_problem}), integrated with $\Delta t = 1/40$
    using the Boris~(\ref{fig:boris_sos-40}), Vay~(\ref{fig:vay_sos-40}),
    and new~(\ref{fig:hc_sos-40}) integrators.}
  \label{fig:surf_sec-40}
\end{figure}

We first use all integrators with a very small time step,
$\Delta t = 1/40$, which corresponds to $80 \pi \approx 250$
time steps per period of the oscillation in $x$ for the given
potential.  The results are shown in Fig.~\ref{fig:surf_sec-40}.
At this temporal resolution, all integrators are seen to give
nested surfaces in this plane, in essence showing that they are
all preserving the invariants of the problem.

\begin{figure}[htbp]
  \centering
  \begin{subfigure}[b]{0.3\textwidth}
    \includegraphics[width=\textwidth]{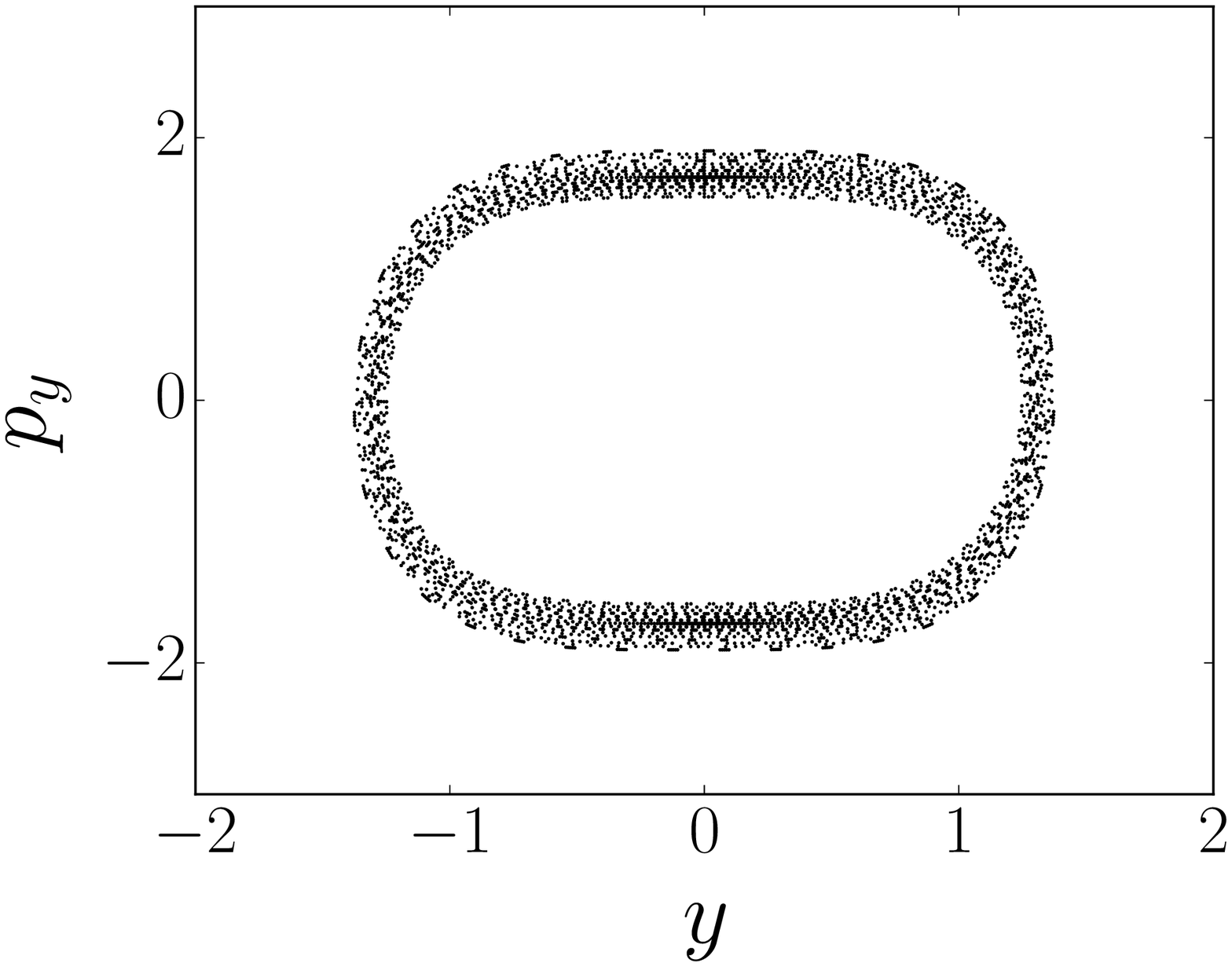}
    \caption{Boris}
    \label{fig:boris_sos-10}
  \end{subfigure}
  \begin{subfigure}[b]{0.3\textwidth}
    \includegraphics[width=\textwidth]{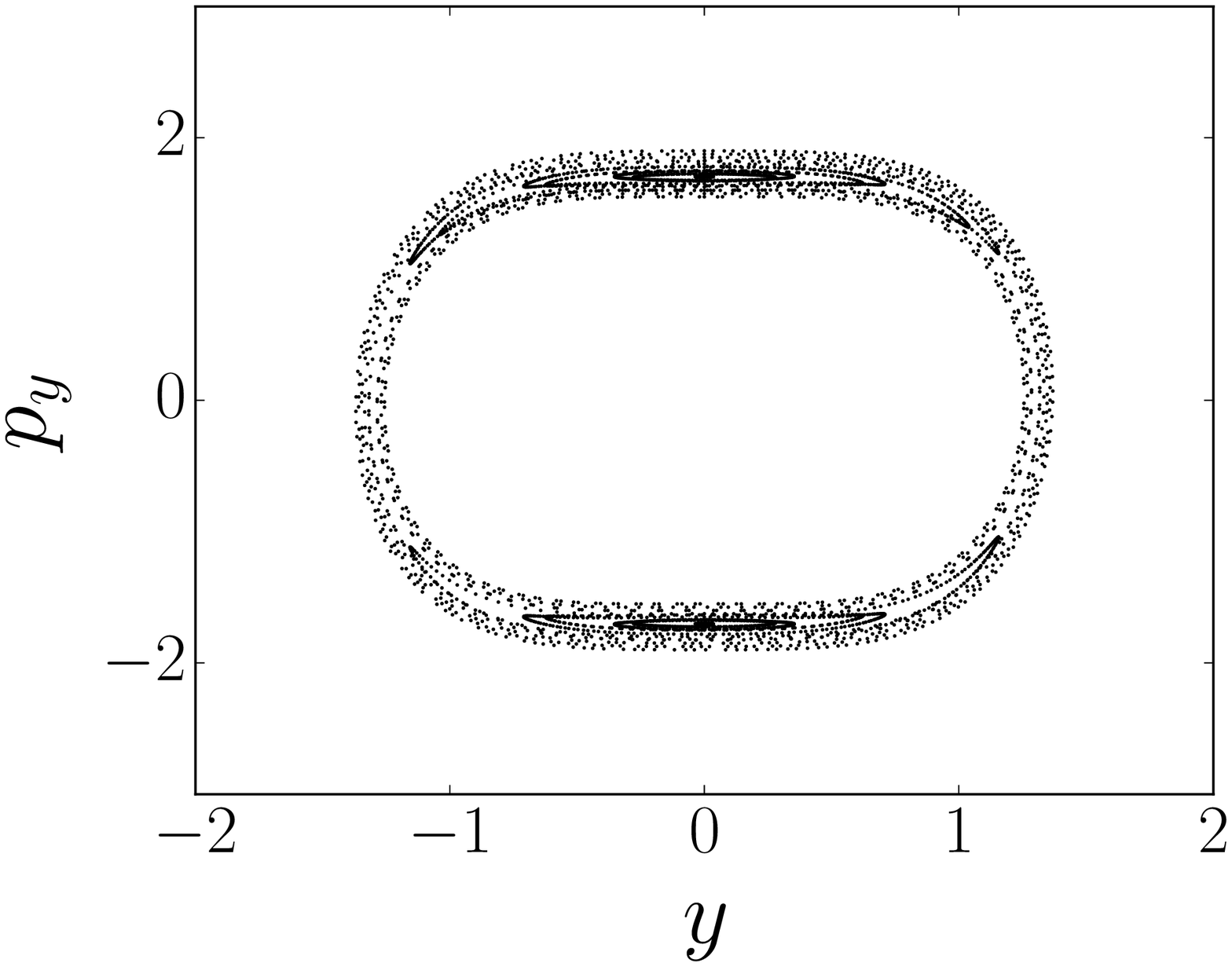}
    \caption{Vay}
    \label{fig:vay_sos-10}
  \end{subfigure}
  \begin{subfigure}[b]{0.3\textwidth}
    \includegraphics[width=\textwidth]{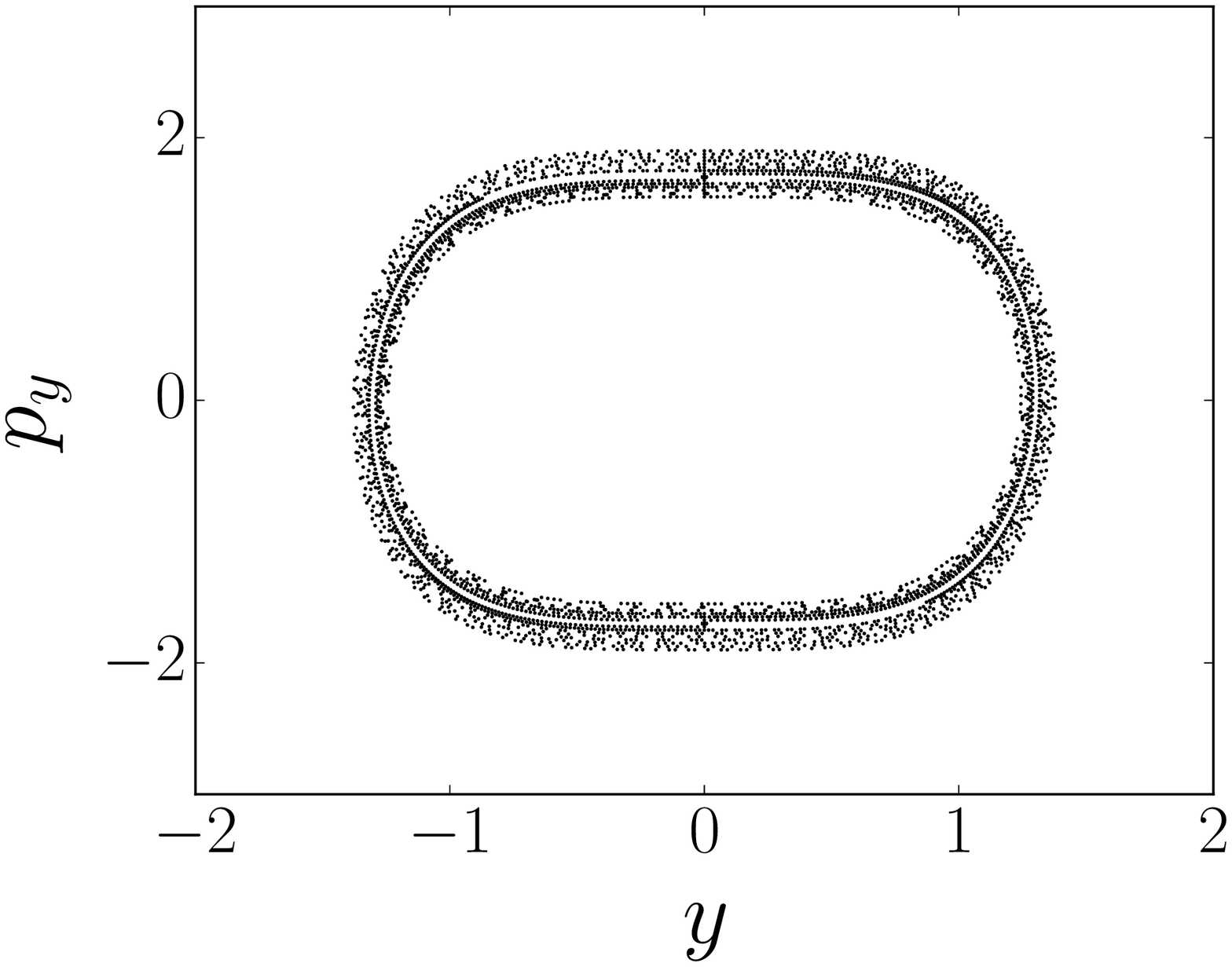}
    \caption{New}
    \label{fig:hc_sos-10}
  \end{subfigure}
  \caption{Poincar\'e surfaces of section for
    system~(\ref{eq:test_problem}), integrated with $\Delta t = 1/10$
    using the Boris~(\ref{fig:boris_sos-10}), Vay~(\ref{fig:vay_sos-10}),
    and new~(\ref{fig:hc_sos-10}) integrators.}
  \label{fig:surf_sec-10}
\end{figure}

Next we use a time step, $\Delta t = 1/10$ that is more typically
used in simulations.  This corresponds to $20 \pi \approx 60$
time steps per period of the oscillation in $x$ for the given
potential.  The results are shown in Fig.~\ref{fig:surf_sec-10}.
At this temporal resolution, the Boris and new integrators continue
to show nested surfaces in this plane, showing that they are accurately
representing the topology of the trajectories.  However, for the
Vay integrator with the initial condition $p_y \approx 1.7$ a resonance
island is seen.  (This is a two-fold degenerate resonance; trajectories
started in the top island do not visit the bottom island.)
Thus, a finite volume of trajectories is trapped in this resonance.
Moreover, the sections for different trajectories are seen to cross
each other.  This is an indication that the other invariants (either
energy or $p_z$) are less well-preserved, as crossing can occur only
if the trajectories are for different values of the other invariants.
Thus, in this example, with a reasonable choice for the time step,
use of the Vay integration leads to unphysical consequences.

\section{Summary and future directions}
\label{summfuture}

We have derived a new integrator for charged particle motion in
arbitrary electromagnetic fields that is both volume-preserving
(like the Boris push) and also correctly computes the $\exb$
drift velocity (like the Vay push).  This new integrator has been
tested numerically and compared with the other integrators.  It
is found not to introduce sizable resonances at reasonable values
for the time step in contrast with the integrator of
\cite{vay2008simulation}.

A number of new directions deserve attention in this area.
What are the consequences of composing integrators for which the
volume alternately grows and shrinks, as it does for the Vay
integrator?  Can one have attractors and repellers in this case?
How does one extend these methods to higher order, as was done
for symplectic integrators by, e.g.,
\cite{yoshida1990construction} and others.  More importantly,
what is gained?  For beam tracking, one needs a spatial version
of the current integrator.  Combining higher-order with spatial
integration could be very powerful for beam tracking.  Finally,
the extension of these concepts to self-consistent simulations,
with higher-order integrators, could be explored.

\section*{Acknowledgements}
This work is supported by DOE/NSF Grant No. DE-SC0012584.
\section*{References}
\label{references}

\bibliography{empush}

\begin{thebibliography}{10}
\expandafter\ifx\csname url\endcsname\relax
  \def\url#1{\texttt{#1}}\fi
\expandafter\ifx\csname urlprefix\endcsname\relax\def\urlprefix{URL }\fi
\expandafter\ifx\csname href\endcsname\relax
  \def\href#1#2{#2} \def\path#1{#1}\fi

\bibitem{boris1970relativistic}
J.~Boris, Relativistic plasma simulation-optimization of a hybrid code, in:
  Proc. Fourth Conf. Num. Sim. Plasmas, Naval Res. Lab, Wash. DC, 1970, pp.
  3--67.

\bibitem{vay2008simulation}
J.-L. Vay, Simulation of beams or plasmas crossing at relativistic velocity,
  Physics of Plasmas (1994-present) 15~(5) (2008) 056701.

\bibitem{cary2016select}
J.~R. Cary, D.~T. Abell, G.~I. Bell, B.~M. Cowan, J.~R. King, D.~Meiser, I.~V.
  Pogorelov, G.~R. Werner, Select advances in computational accelerator
  physics, IEEE Transactions on Nuclear Science 63~(2) (2016) 823--841.

\bibitem{birdsall2004plasma}
C.~K. Birdsall, A.~B. Langdon, Plasma physics via computer simulation, CRC
  Press, 2004.

\bibitem{hockney1988computer}
R.~W. Hockney, J.~W. Eastwood, Computer simulation using particles, CRC Press,
  1988.

\bibitem{nieter2004vorpal}
C.~Nieter, J.~R. Cary, Vorpal: a versatile plasma simulation code, Journal of
  Computational Physics 196~(2) (2004) 448--473.

\bibitem{goldstein2014classical}
H.~Goldstein, C.~P. Poole, J.~L. Safko, Classical Mechanics: Pearson New
  International Edition, Pearson Higher Ed, 2014.

\bibitem{ruth1983canonical}
R.~D. Ruth, et~al., A canonical integration technique, IEEE Trans. Nucl. Sci
  30~(4) (1983) 2669--2671.

\bibitem{feng1986difference}
K.~Feng, Difference schemes for hamiltonian formalism and symplectic geometry,
  Journal of Computational Mathematics 4~(3) (1986) 279--289.

\bibitem{forest1990fourth}
E.~Forest, R.~D. Ruth, Fourth-order symplectic integration, Physica D:
  Nonlinear Phenomena 43~(1) (1990) 105--117.

\bibitem{candy1991symplectic}
J.~Candy, W.~Rozmus, A symplectic integration algorithm for separable
  hamiltonian functions, Journal of Computational Physics 92~(1) (1991)
  230--256.

\bibitem{yoshida1990construction}
H.~Yoshida, Construction of higher order symplectic integrators, Physics
  Letters A 150~(5) (1990) 262--268.

\bibitem{cary1993explicit}
J.~Cary, I.~Doxas, An explicit symplectic integration scheme for plasma
  simulations, Journal of Computational Physics 107~(1) (1993) 98--104.

\bibitem{qin2013boris}
H.~Qin, S.~Zhang, J.~Xiao, J.~Liu, Y.~Sun, W.~M. Tang, Why is boris algorithm
  so good?, Physics of Plasmas (1994-present) 20~(8) (2013) 084503.

\bibitem{zhang2015volume}
R.~Zhang, J.~Liu, H.~Qin, Y.~Wang, Y.~He, Y.~Sun, Volume-preserving algorithm
  for secular relativistic dynamics of charged particles, Physics of Plasmas
  (1994-present) 22~(4) (2015) 044501.

\bibitem{stoltz2002efficiency}
P.~Stoltz, J.~Cary, G.~Penn, J.~Wurtele, Efficiency of a boris-like integration
  scheme with spatial stepping, Physical Review Special Topics-Accelerators and
  Beams 5~(9) (2002) 094001.

\end{thebibliography}

\end{document}